\documentclass[seceq]{ptptex}

\usepackage{graphicx}

\def\bi{\bibitem}
\def\la{\langle}\def\ra{\rangle}
\def\be{\begin{eqnarray}}\def\ba{\begin{eqnarray}}
\def\ee{\end{eqnarray}}\def\ea{\end{eqnarray}}
\def\ben{\begin{enumerate}}\def\bitem{\begin{itemize}}
\def\een{\end{enumerate}}\def\eitem{\end{itemize}}
\def\Tr{\rm Tr}
\def\pr{Phys. Rev.}\def\prl{Phys. Rev. Lett.}
\def\np{ Nucl. Phys.}\def\pl{Phys. Lett.}
\def\lsim{\mathrel{\rlap{\lower3pt\hbox{\hskip1pt$\sim$}}
     \raise1pt\hbox{$<$}}} 
\def\gsim{\mathrel{\rlap{\lower3pt\hbox{\hskip1pt$\sim$}}
     \raise1pt\hbox{$>$}}}

\title{
Hidden Local Symmetry and the Vector Manifestation \\ of Chiral
Symmetry in Hot and/or Dense Matter\footnote{Based in part on talks
given at {\it The 1st Asian Triangle Heavy Ion Conference}, Yonsei
University, Seoul, Korea, 29 June - 1 July 2006 and at {\it Yukawa
International Seminar on New Frontiers in QCD}, Yukawa Institute,
Kyoto, Japan, 3 - 8 December 2006.}
 }

\author{
Mannque \textsc{Rho}$^{1,}$\footnote{ e-mail address:
mannque.rho@cea.fr}}

\inst{

$^1$ Service de Physique Th\'eorique, CEA Saclay, 91191
Gif-sur-Yvette, France}

\abst{The emergence and importance of hidden local symmetry in the
structure of hadrons under extreme conditions is discussed. The
topics covered are the potentially important role played by an
infinite tower of vector mesons encoded in holographic dual QCD (or
AdS/QCD) in chiral dynamics of mesons and baryons, in the vector
dominance and its violation in EW response functions and the
presence of the vector manifestation fixed point and its influence
on the properties of hadrons in hot temperature (i.e., in
relativistic heavy ion collisions) and in dense matter (i.e., in
compact stars). Also discussed are the natural emergence of
instantons/skyrmions from the infinite tower of vector mesons
coupled to pions in AdS/QCD and their (speculated) role in chiral
restoration at high density. }

\begin{document}

\maketitle


\section{Introduction}
In the current understanding of quantum chromodynamics (QCD), nearly
all of the masses of low-lying hadrons (e.g., $\sim 98$\% of the
proton mass) come from spontaneous breaking of chiral symmetry. A
hot issue raised currently by hadron/nuclear physicists is: How to
un-break the broken symmetry in terrestrial laboratories or in
compact stellar systems and figure out how the mass is generated to
start with.  I would like to address this issue in this talk using
the framework Gerry Brown and I have been developing since some
time, specifically, ideas based on hidden local symmetry, the
``vector manifestation" of chiral symmetry and Brown-Rho scaling.

The above question belongs broadly to a class of issues connected to
the origin of mass. In a glaring departure from molecules, atoms and
nuclei whose masses are nearly fully, say, more than 99\%, accounted
for by their ``elementary" constituents, the bulk of the mass of the
nucleon is not given by the masses of its constituents, namely, the
quarks. Thus this sets the beginning of a series of mysteries
related to the question ``where does the mass come from?"

The clue to the origin of the mass of quarks and leptons, the truly
fundamental issue, will probably be revealed in the coming years,
perhaps at LHC/CERN. What we are concerned with here is the issue --
perhaps less fundamental -- we can address with what we already have
at our disposal and will have in the near future and deals with
strongly interacting matter when it is heated to several hundred MeV
in temperature or is compressed to several times the density of the
nuclear matter.
\section{The Origin of Hadron Mass}
Let us take the proton as a typical hadron. The argument goes in a
similar way for, say, the $\rho$ meson mass which is currently a hot
topic of experimental efforts. The simplest way to proceed is to
think in terms of constituent quarks whose masses are dynamically
generated by the complex vacuum.

In QCD, the proton is made up of three light (``chiral") quarks, the
total mass of which is tiny, a few MeV, compared with the proton
mass which is $\sim$ 1000 MeV. It is believed - and the evidence is
strong - that most of the proton mass comes from the ``spontaneous"
breaking of chiral symmetry by the vacuum, with the order parameter
given by the quark condensate $\la\bar{q}q\ra$. The mass so
generated must therefore be connected intimately to the quark
condensate. So in the chiral limit, one should be able to write the
hadron mass as a function of the condensate as
 \be
m=F(\la\bar{q}q\ra).
 \ee
As an order parameter, the condensate $\la\bar{q}q\ra$ will be
assumed to go to zero when chiral symmetry is restored. It may go to
zero either in a smooth way or in a discontinuous way when the
system is driven by the external condition - temperature or density
- to the critical point. Now even if the condensate goes to zero
smoothly, the $F$ which could be a complicated function of the
condensate may go to zero in a discontinuous way. At present, there
is no known analytical way to settle this issue using QCD proper. So
how $F$ behaves as the critical point is reached is not known. But
it seems very natural to expect that
 \be
F(\la\bar{q}q\ra)\rightarrow 0\ \ {\rm as}\ \
\la\bar{q}q\ra)\rightarrow 0.
 \ee
This is basically the basis of Brown-Rho scaling~\cite{BR:91} and is
supported by Harada-Yamawaki's hidden local symmetry theory with the
``vector manifestation" fixed point~\cite{HY:PR}. We will call this
Harada-Yamawaki theory ``HLS/VM" for short.

We should however mention that this is not the only possibility. In
fact, certain models can admit a non-zero $F(0)$, hence a non-zero
mass, and still preserve chiral symmetry and allow a
Goldstone-Wigner phase transition. For instance, one can have a
linear sigma model with parity-doublet fermion
fields~\cite{parity-doublet}. In this model, the fermion mass can be
non-zero in both the Goldstone and Wigner phases. The fermions can
be baryons or constituent (quasi) quarks. In the former case, one
can have massless mesons while having massive baryons at the
critical point. In the latter case, both mesons and baryons as
(weakly) bound states of quasiquarks could remain massive at the
critical point with the chiral symmetry preserved.  At present,
neither theory nor experiment can rule out this scenario. However we
find it unnatural within the HLS framework we are adopting. We shall
not pursue this alternative scenario.

\section{Hidden Local Symmetry (HLS)}
The well-established perturbative QCD cannot access the highly
nonperturbative and strong-coupling regime at low energy we are
concerned with. The only tool available at present is effective
theories that meet Weinberg's ``folk theorem"~\cite{weinberg-folk}
on effective field theories. There is some help from lattice
calculations but at the present stage, not much on what happens to
light-quark hadrons, e.g., nucleons, vector mesons, pions etc. when
they are immersed in hot and/or dens medium.

The question we are interested in answering is: If the hadron mass
vanishes at the phase transition, how can one ``see" it or what is
the appropriate tool for it?

In what follows, we shall mostly deal with the chiral limit. Our
principal theme in addressing the above question is that {\it the
most important ingredient in an effective field theory that enables
one to probe the regime where the effective mass of the vector meson
can drop to that of the pion mass is hidden local symmetry.} Our
theme is anchored on the argument by Harada and
Yamawaki~\cite{HY:PR} that without hidden local symmetry, there is
no consistent {\it as well as} simple way to allow the vector mass
become as light as the pion mass. We suggest that the lack of hidden
gauge symmetry is the reason why phenomenological models often used
in the literature fail to observe dropping masses in hot/dense
matter.
\subsection{HLS as emergent gauge symmetry}
At very low energy $E\ll \Lambda_\chi$ where $\Lambda_\chi\approx
4\pi f_\pi\sim 1$ GeV, the only relevant degrees of freedom are the
pions, that is, Goldstone bosons, emerging from the spontaneous
breaking of chiral symmetry $SU(N_f)_L\times SU(N_f)_R$ to
$SU(N_f)_{L+R}$. The dynamics involving the pion field,
$U=exp(2i\pi/F_\pi)$~\footnote{Notations: $\pi=T^a\pi^a$ and $\Tr
(T^a T^b)=\frac 12 \delta^{ab}$.} is encoded in a chiral Lagrangian
expanded in derivatives, the leading term of which is given by
low-energy theorems. Now we make the most obvious and ``trivial"
observation that one can always write the $U$ field in a product
form if one is willing to introduce a redundant field. Define the
L/R chiral fields with the redundant field $\sigma$
 \be
\xi_{L,R}=e^{\mp i\pi/F_\pi}e^{i\sigma/F_\sigma}\label{chifields}
 \ee
$U$ can then be rewritten as
 \be
U=\xi_L^\dagger\xi_R.
 \ee
In this form, we unearth a local symmetry
 \be \xi_{L,R}\rightarrow h(x)\xi_{L,R}
 \ee
with $h(x)\in SU(N_f)_V$. As it stands, we have not done anything
new, so there is no new physics here. However if we elevate the
local symmetry to a local gauge symmetry by introducing a gauge
field -- which we will denote by $V_\mu \in SU(N_f)_V$ -- and endow
it with a kinetic energy term, then it becomes quite a different
story. First of all, this procedure allows one to go up
systematically in some expansion scheme (such as chiral expansion)
in energy from the low energy scale where the pionic chiral
Lagrangian is applicable to a scale at which new degrees of freedom
set in~\cite{georgi-idea}, thereby circumventing the breakdown of
the pionic chiral theory and going beyond to the next energy scale.
In our case, the scale is brought above the mass of the vector
mesons $\rho$, $\omega$, $a_1$ etc. Furthermore. what is more
important for our case is that thanks to local gauge symmetry, one
can do a systematic chiral perturbation calculation with the vector
mesons put on the same scale as the pions and access the regime
where the vector mass is comparable to the pion mass.

The hidden local symmetry theory of Harada and Yamawaki\cite{HY:PR}
-- which is based on the earlier work of Bando et
al.~\cite{bandoetal} -- was constructed in this way with the vector
mesons $\rho$ and $\omega$ (which we label as $V_1$) as the hidden
local fields together with the Goldstone pions.

This strategy of introducing a hidden gauge field can be extended to
as many gauge fields as required by the energy scales one wants to
deal with. There is however one important caveat here. Introducing
gauge degrees of freedom in this way as a way of going up in energy
scales does not necessarily lead to a unique gauge
theory~\cite{weinberg-largeNc}. Different constructions could give
rise to different gauge structures at higher energies. One way that
is both simple and phenomenologically appealing is the linear
``moose" construction~\cite{georgi-idea}. A version of this class of
construction in which an infinite tower of gauge fields are
incorporated was proposed as a ``dimensionally deconstructed
QCD"~\cite{son-stephanov1}. In order to make the construction
correctly represent nature up to a given scale, the theory has to be
matched to a fundamental theory and ``ultraviolet-completed" to go
beyond.

As we will elaborate further later, if one wants to study what
happens to the vector mesons in medium, this strategy is definitely
needed. On a more fundamental level, one can view this as a generic
phenomenon of the ``emergence" of local gauge degrees of freedom.
Examples are numerous, e.g., emergent gravity, emergent space-time,
spin-charge separation in high-T superconductivity
etc~\cite{horowitz-polchinski}.
\subsection{HLS from string theory}
A novel recent development in both string theory and hadronic
physics is that HLS $descends$ naturally via AdS/CFT duality from
string theory to holographic QCD. This is a top-down approach to
hidden local symmetry theory of QCD. One such theory which astutely
implements the spontaneous breaking of chiral $SU(N_f)\times
SU(N_f)$ symmetry was constructed by Sakai and Sugimoto~\cite{SS}.
The key idea in this approach is that the strongly-coupled and hence
highly non-perturbative aspect of QCD in four dimensions in the
limits $\lambda\equiv g^2_{YM}N_c\rightarrow \infty$ (``'t Hooft
limit") and $N_c\rightarrow \infty$ (``large $N_c$ limit") can be
approximated by a readily calculable weakly-coupled gravity solution
in five dimensions. Since a dimensional reduction is involved, there
is a Kaluza-Klein scale $M_{KK}$ that sets the energy scale of the
effective theory. What results is a pure Yang-Mills theory in five
dimensions dual to strongly-coupled QCD which when reduced to four
dimensions can be cast in terms of an infinite tower of local vector
fields coupled gauge invariantly to the pions. Thus as in the
emergent case, the same type of hidden local symmetry (with an
infinite tower of vector mesons and the pions) arises from top down.
Remarkably this theory describes the meson sector~\cite{SS} as well
as the baryon sector~\cite{HRYY-PRL,HSSY} quite well.

Perhaps the most important outcome of the development is the
appearance of baryons in the theory. The hidden local symmetry
theory (with the infinite tower which we will label as $V_\infty$
whenever unambiguous) is the full theory of hadrons at the scale
defined by the KK mass $M_{KK}$. Since there are no explicit baryons
in the theory,  baryons must arise through topology, namely,
solitons. In five dimensions, the soliton is an instanton but
reduced to four dimensions where the baryon lives, it is a
skyrmion~\cite{skyrme}. What makes this skyrmion different from the
skyrmion in the Skyrme model~\cite{skyrme} is that the soliton
involves the infinite tower of vector mesons in addition to the
pions encapsulated in an instanton. The geometry in five dimensions
represents the dynamics of the infinite tower of vector mesons. The
bulk theory is weak-coupling and manageable in the limit
$\lambda\rightarrow \infty$ and $N_c\rightarrow \infty$ and provides
baryon chiral dynamics anchored on hidden gauge structure.

There is a simple prediction of the theory that can be checked
against experiments. Viewed in terms of the instanton, the nucleon
axial coupling $g_A$ is independent of $\lambda$ and the KK scale
$M_{KK}$, so it is independent of the pion decay constant. One can
also calculate in the same limit the anomalous magnetic moment of
the nucleon $\mu_{an}=\mu_p -\mu_n -1$ in nuclear magneton. Both are
linear in $N_c$. The ratio $R=g_A/\mu_{an}$ is therefore independent
of $N_c$ in the 't Hooft and large $N_c$ limit. The instanton
calculation~\cite{HRYY-PRL} is found to give $
 g_A\approx 0.69\frac{N_c}{3} +{\cal O}(N_c^0)$
and $\mu_{an}\approx 2.16\frac{m_N}{M_{KK}}\frac{N_C}{3}+{\cal
O}(N_c^0)$. With the ${\cal O}(1)$ term ignored for the moment, the
predicted ratio for $M_{KK}\approx m_N$ as required in the meson
sector~\cite{SS} comes out to be $R\approx 0.32$. The experimental
values are $g_A=1.26$ and $\mu_{an}=3.7$, so the empirical ratio
$R=0.34$ gives a support to the prediction.

One can go one step further and estimate the ${\cal O}(1)$
correction with no sweat. In fact, one can argue that $1/N_c$
corrections from non-planar loop contributions come only at ${\cal
O}(1/N_c)$ and higher. The ${\cal O}(1)$ corrections are common in
both $g_A$ and $\mu_{an}$ and they arise in a simple book-keeping of
the spin-flavor operators involved in the hedgehog configuration
that is totally independent of dynamics. The leading correction is
given simply by the replacement $N_c\rightarrow N_c+2$. Thus the
final prediction for $N_c=3$ is $g_A \approx 1.15$ and
$\mu_{an}\approx 3.6$ in close agreement with the experimental
values~\footnote{The instanton baryon gives an additional correction
of ${\cal O}(1/(\lambda N_c))$ to $g_A$ which for the value of
$\lambda N_c$ fixed in the meson sector by, say, $f_\pi$ etc, comes
out to be $g_{A_{min}}\approx 0.15$, so the total $g_A\approx 1.31$
comes even closer to the experiment. The nature of approximations
involved in this calculation appears to be quite similar to the
quenched approximation in lattice QCD which also predicts $g_A$
close to the experimental value~\cite{quenchedgA}.}

Another remarkable prediction of the instanton baryon
description~\cite{HRYY-long} is that the nucleon EM form factors are
completely vector-dominated just as the pionic form factor is. The
photon can couple directly to the soliton which ia an extended
baryon but by a suitable field re-definition, one can ``eliminate"
the direct photon coupling and express the form factors entirely
vector-dominated with the infinite tower of vector mesons entering
into the formula. This result may offer a precise meaning to what
was previously considered as a ``intrinsic core" attributed to a
quark-gluon structure in the two-component model of the nucleon
structure~\cite{2comp-NFF,petronzio}. It furthermore brings a
basically new aspect to the solitonic structure of the baryon which
is now governed by the vector mesons, not by the Goldstone (pion)
field as in the skyrmion with the Skyrme model. What happens to this
instanton baryon in hot/dense medium is totally unknown.

At the present stage of our understanding, the holographic approach
can make predictions only in the large $\lambda$ and $N_c$ limit,
restricted to the zero temperature and matter-free environment. In
studying hadron properties in medium, however, it is clear that one
has to be able to calculate $1/N_c$ corrections since the hadron
masses are locked to the quark condensate as we argued above and in
the large $N_c$ limit, the quark condensate is known to be
temperature-independent~\cite{large-N-condensate}. At present, one
does not know how to compute higher-order $1/N_c$ terms in the bulk
sector. Clearly further progress in this direction is needed for the
approach to describe what happens in the vicinity of the chiral
restoration, a matter of prime importance in the field.
\section{Vector Manifestation}
\subsection{HLS \`a la Harada and Yamawaki}
Since one cannot yet adequately exploit the dynamics of the infinite
tower in confronting nature, we will rely on the HLS theory of
Harada and Yamawaki (HY)~\cite{HY:PR} which involves only the lowest
members $\rho$ (and $\omega$) of the infinite tower. Let us call it
HLS$_1$ in contrast to the infinite-tower HLS theory of Sakai and
Sugmoto which we will call HLS$_\infty$. The HLS$_1$ theory can be
interpreted as a truncated version of HLS$_\infty$ in the following
sense. One picks the scale $\Lambda_M$ as a matching scale,
integrates out all the members of the tower except the lowest member
$V_1$ lying below $\Lambda_M$, writes HLS Lagrangian with the $V_1$
and the pion and Wilsonian-matches at $\Lambda_M$ the HLS
correlators to the QCD correlators. One can obviously include in the
theory other vector mesons as for instance the axial vector meson
$a_1$ etc. This gives a $bare$ Lagrangian whose parameters are given
by QCD variables such as the strong coupling constant $\alpha_s$,
the quark condensate $\la\bar{q}q\ra$, the gluon condensate $\la
G^2_{\mu\nu}\ra$ etc. Quantum calculations are done with the bare
HLS Lagrangian so determined by renormalization group equations.

An important point to note here: {\it Since the condensates in the
QCD sector are background-dependent, that is, dependent on
temperature, density etc,  the bare Lagrangian HLS$_1$ is endowed
with ``intrinsic (background) dependence."}
\subsection{The fixed point of HLS$_1$}
In the chiral limit and with $V_1=\rho$, there are three parameters
in the HLS$_1$ Lagrangian, $g$, $F_\pi$ and $a$ where $g$ is the
gauge coupling constant of HLS$_1$, $F_\pi$ is the parametric pion
decay constant~\footnote{The physical pion decay constant will be
written as $f_\pi$.} and $a=(F_\sigma/F_\pi)^2$ figuring in the
chiral fields (\ref{chifields}). Harada and Yamawaki showed that the
RGEs for these parameters have the fixed point consistent with
QCD~\footnote{There are other fixed points in the theory but they
belong to a different universality class than that of QCD.}
 \be
g^*=0, \ a^*=1.\label{fixed}
 \ee
This was explicitly shown to one-loop order but one can readily
convince oneself that it should hold to all loop orders. This fixed
point dictates how the hadronic system makes a phase transition from
the chiral symmetry broken phase to the restored phase. {\it The
fixed point (\ref{fixed}) -- referred to as ``vector manifestation
(VM) fixed point" -- is reached when the quark condensate vanishes.}

In HLS$_1$ (as well as HLS$_\infty$) theory, the vector meson masses
are generated by the Higgs mechanism, with the scalar $\sigma$ eaten
by the vector mesons. The mass formula therefore is
 \be
m_V^2=g^2 F_\sigma^2=a g^2 F_\pi^2.
 \ee
It has been shown that near the critical point, either $T_c$ or
$n_c$, the coupling $g$ goes to zero proportionally to
$\la\bar{q}q\ra$. Therefore at the chiral restoration, the
parametric mass behaves
 \be
m_V\sim g\sim \la\bar{q}q\ra\rightarrow 0.
 \ee
This is consistent with Brown-Rho scaling. The physical mass of
$V_1$ also vanishes at the critical point.

The parametric pion decay constant does not exhibit any special
feature as $\la\bar{q}q\ra\rightarrow 0$. However the physical pion
decay constant $f_\pi=F_\pi +\delta f_\pi$ must go to zero as
dictated by the low-energy theorem.
\subsection{Vector dominance violation}
An important consequence for the problem we are concerned with of
the VM fixed point is that the vector dominance in the EM form
factors of hadrons is drastically affected. To see this, we look at
the pion EM form factor. In the HLS$_1$ theory, the photon couples
to the point-like pion with a coefficient $(1-a/2)$ and through the
$V_1=\rho$ meson with a coefficient $a/2$. In matter-free space,
$a\approx 2$, so the form factor is vector-dominated with the direct
coupling vanishing. What Harada and Yamawaki found was that the
$a=2$ point sitting in free space is purely $accidental$, with $a$
lying on a RG trajectory that does not contain $a=2$. Thus if the
system is slightly perturbed, $a$ quickly flows from 2 toward 1. At
the VM fixed point, therefore the standard vector dominance is
$strongly$ violated~\footnote{The violation of the standard vector
dominance -- phrased in terms of the lowest members $(\rho, \omega)$
of the tower -- can be interpreted as the intervention of the higher
members of the infinite tower. It is not clear what the ``fixed
point" $a=1$ in HLS/VM means in the infinite-tower structure of the
vector dominance found in AdS/QCD~\cite{HRYY-long}. It is also not
clear why $a$ cannot go below 1 at some point in this model.}.
\section{``Seeing" the Dropping Mass}
\subsection{On-shell probes}
There is a flurry of activities, both in theory and experiment, to
``see" evidence for the manifestation of chiral symmetry in the
behavior of masses and coupling constants in hot and/or dense
medium. One would first like to observe it at a density and/or
temperature readily available in laboratories before reaching the
critical point and then ultimately go to the critical point. The
former is looked for in precursor phenomena to the chiral phase
transition. The idea is to create a hot/dense environment in which
the produced particles propagate and look for signals that take a
``snap-shot" of the particle propagation in the medium that is as
free as possible from subsequent interactions with the medium.

Among the weakly interacting probes, the most frequently used is the
electromagnetic one. There are two classes of processes that have
been studied: One mediated by the $normal$ component of the chiral
Lagrangian and the other by the $anomalous$ component.
 \ben
\item The lepton pair production via virtual photon
 \be
V\rightarrow \gamma^*\rightarrow l^+ l^-\label{V}
 \ee
where $V=\rho, \ \omega$ and $l=e, \ \mu$ is governed by the
$normal$ component of the HLS$_1$ Lagrangian. Most of the past
efforts to unravel chiral dynamics of hadrons in medium were
directed to this process. Several heavy-ion experiments have been
dedicated at CERN-SPS to expose $directly$ in-medium properties of
the $\rho$ meson in hot and dense medium. The most recent and
comprehensive review on the current situation regarding this class
of experiments can be found in the articles by van Hees and
Rapp~\cite{rapp}. We will argue below that within the HLS/VM theory
that we believe is suited to the problem, the lepton pairs in
heavy-ion processes are {\it neither efficient nor specific} for
exposing in-medium properties of the $\rho$ meson bearing on chiral
symmetry, in particular, on the ``dropping mass" effect.
\item The other class involves processes that are mediated by
the $anomalous$ Wess-Zumino term in chiral Lagrangians. The process
studied in this category~\cite{metag} involves the in-medium
$\omega$ meson in
 \be
\gamma +A\rightarrow \omega +X\rightarrow \pi^0\gamma +
X^\prime.\label{Metag}
 \ee
The coupling $\omega\pi^0\gamma$ is significant in that it arises
from a chiral anomaly and is expected to behave differently in
medium from that of the normal process (\ref{V}) as certain
anomalous processes (e.g., those involving triangle diagrams) are.
 \een
\subsection{Off-shell probes}
There are indirect probes that indicate how hadron masses and
coupling constants behave in dense medium. A recent discussion on
this matter can be found in the review by Brown et al~\cite{BHLR06}.
Most of the relevant arguments given there have been developed in a
series of articles that date way back to early 1990's~\cite{BR:PR}.

If one looks for a signal at zero temperature but at a matter
density near that of the normal nuclear matter, then the mass
scaling
 \be
\Phi (n)=m^*/m \equiv m(n)/m(0)
 \ee
is a $parameter$ in an effective theory because what one is dealing
with here is an off-shell quantity. It can however be related to the
quark condensate, albeit indirectly, and hence could signal for a
chiral symmetry manifestation. In order to ``see" the effect in an
unambiguous way, one would have to formulate the effective theory in
a way consistent with chiral symmetry {\it for all scales} involved
in the process. There are a vast variety of such formulations in the
literature and a direct comparison with them is not our aim here. As
an indication of the subtlety involved in the problem, let us take
an exemplary case discussed by us\cite{BR:PR}, i.e., the effective
chiral Lagrangian in the mean field that can be recast in Landau
Fermi-liquid theory for nuclear matter. There the scaling $\Phi (n)$
for $n\approx n_0$ (where $n_0$ is the nuclear matter density) can
be related to the Landau parameter
$\tilde{F}_1$~\footnote{$\tilde{F}$ is related to the Landau
parameter $F$ by $\tilde{F}=\frac{m_N}{m_L} F$ where $m_L$ is the
Landau quasiparticle mass of the nucleon.} as\footnote{It should be
stressed that this relation holds only in the vicinity of the
nuclear saturation density, i.e., the Fermi-liquid fixed point and
cannot be extrapolated to too high a density.}
 \be
\tilde{F}_1 -\tilde{F}_1^\pi\approx 3(1-1/\Phi)\label{LandauF}
 \ee
where $\tilde{F}_1^\pi$ is the contribution to the Landau parameter
from the pion exchange which is completely fixed by chiral dynamics.
One can get $\tilde{F}_1$ from the anomalous gyromagnetic ratio of
the proton $ \delta g_l$ measured experimentally by using the Migdal
formula that relates the two quantities~\cite{Migdal}. Using
(\ref{LandauF}), one finds~\cite{Friman-Rho}
 \be \Phi (n_0)\approx 0.78.\label{Phivalue}
 \ee
In HLS$_\infty$ theory, the Landau parameter $F_1$ must be receiving
contributions from all of the $V_\infty$ in the $\omega$ channel.
However assuming that the lowest $\omega$ dominates, we can
interpret (\ref{Phivalue}) as
 \be
m_\omega (n_0)/m_\omega\approx 0.78.
 \ee
While this is valid only near the nuclear matter density for the
``$\omega$" meson in the average sense, it is comparable -- as
predicted by Brown-Rho scale valid up to $\sim n_0$ -- to the ratio
$f_\pi (n_0)/f_\pi(0)\approx 0.80\approx \sqrt{\frac{\la\bar{q}q\ra
(n_0)}{\la\bar{q}q\ra}}$ found in deeply bound pionic atoms.

Other indirect indications for the dropping masses,  such as for
example the reduction of the nuclear tensor force etc., are
summarized in the reviews~\cite{BR:PR}. What may be a bit surprising
is that nuclear physics has fared so successfully without ever
invoking Brown-Rho scaling even though the vacuum change due to
density is expected. As suggested in \cite{BR:PR}, this may be
understood in terms of a  web of intricate ``dualities" at low
energy known variously as ``quark-hadron continuity," ``Cheshire Cat
principle" etc., concepts that are yet to be put on a solid basis.

The import of the discussion in this subsection is that it
illustrates how chiral symmetry property which is a basic property
of QCD  and Fermi-liquid property which is a mundane many-body
effect can be compounded in $ordinary$ nuclear processes.
\subsection{What the dileptons see}
Experimentalists in heavy ion physics purport to extract an
in-medium ``spectral function" of a vector meson quantum number,
say, $\rho$ or $\omega$ as a function of invariant mass. This
quantity is intended to tell us how the properties of vector mesons
change when they are embedded in medium. An ``ideal" snapshot for
this is thought to be the dileptons mentioned above. In order to
expose the effect of dense and/or hot medium, one tries to subtract
all possible ``trivial effects" that take place in zero-temperature,
zero-density environment (e.g., ``cocktail events"). Whether this
can be done in a fully consistent way is not clear. The discussion
given in the preceding subsection indicates that the distinction
between what is trivial and what is non-trivial in the context of
chiral symmetry can be highly problematic {\it unless one can probe
exclusively the vicinity of the VM fixed point.} Let us suppose for
the sake of discussion that $all$ such trivial effects can be taken
out of the given experimental results and theorists are given what
we shall call in-medium ``ESF" (experimental spectral function). The
question is: {\it Can the ESF so obtained be used to verify or
falsify Brown-Rho scaling or equivalently see evidence for partial
or complete chiral symmetry restoration?}

As stressed above, to properly address this question, it would be
necessary to have at one's disposal one complete self-consistent
theoretical framework in which calculations can be done for all
processes involved, avoiding mixing different models for different
steps. Furthermore one should express the spectral function in terms
of variables that track the chiral property of the system in terms
of the order parameter, i.e., the quark condensate.

HLS$_1$ theory provides one such framework, with its local gauge
invariance enabling one to access the vector fields whose mass scale
can drop as low as that of the pion near the critical point.

The theoretical ingredients necessary within the framework of HY's
HLS$_1$ theory were spelled out in two unpublished
articles~\cite{BR:NA60}. They are (1) the intrinsic background
(temperature, density) dependence demanded by matching to
QCD~\cite{HY:PR}, (2) the violation of the vector dominance in the
pion EM form factor in hot and/or dense medium that results from the
vector manifestation of chiral symmetry in the HLS/VM
theory~\cite{HS:VD} and (3) many-body correlations generated by the
presence of the Fermi surface, which may be considered as a quantum
critical phenomenon~\cite{BHLR06}.

In order to implement all three ingredients in a consistent way,
baryonic degrees of freedom are mandatory. In HLS$_\infty$ theory,
they are skyrmions in an infinite tower of vector mesons if viewed
in four dimensions. No such formulation exists at the
moment.~\footnote{Initial attempts in this direction using the
Skyrme model were made by Park et al.~\cite{BYP}. But without a
proper account of the vector degrees of freedom, the result remains
largely incomplete.} There is no reason why such a formulation
cannot be made, so we shall assume that we have a spectral function
calculated in that formulation as function of temperature/density
and kinematics, and call it ``TSF" (theoretical spectral function).
Unfortunately, having such a TSF is not enough to directly confront
the ESF. One also has to know precisely what the conditions with
which the measurement of the ESF is made are. The heavy-ion
experiment typically involves summing over the dilepton emissions as
the system evolves in temperature and density as it expands. Many
subtle and complex effects, such as for instance the ``memory
effects" pointed out in \cite{C-Greiner}, will have to be all taken
into account. These are formidable complications, one of the major
obstructions to meaningfully confronting HLS/VM with experimental
data. It is not clear that the theorists have the full control on
all these.

Up to date, no one has succeeded in constructing a spectral function
as a function of temperature and density that takes fully into
account the three ingredients listed above. Presently available is a
partial spectral function obtained by Harada and
Sasaki~\cite{HS:spectral} in which only the temperature effects in
the ingredients (1) and (2) were taken into account. Lacking the
crucial density dependence and also the ingredient (3), one cannot
confront the Harada-Sasaki spectral function with the experimental
data even if one had at one's disposal a reliable evolution code --
which one does not -- and a reliable control on the experimental
conditions. Even so, the analysis of Harada and Sasaki gives a clear
indication as to what the dileptons might be ``seeing" in hot and
dense medium. First of all, mass shifts in the TSF cannot occur
without the intrinsic background dependence (IBD) taken into
account: Hadronic (thermal and dense) quantum (loop) corrections can
broaden the spectra but do not shift the invariant mass. Next as a
consequence of the IBD, a finite temperature and density take the
parameter $a$ away from the vector dominance point $a=2$ which is
sitting on an unstable trajectory and makes the VD violated,
maximally at the VM fixed point with $a=1$. Lepton pairs will be
produced more or less equally from the pions by direct coupling and
through vector mesons, thereby modifying the spectral structure from
that in which vector dominance is assumed, a procedure which is
practiced in the field. The HLS/VM predicts unequivocally that the
vector mass is shifted and the width of the vector meson gets
narrower in the close vicinity of the VM fixed point. However as
measured by the lepton pairs that come from both direct pions and
vector mesons in the system that evolves both in temperature and
density, the width will be smeared while the strength will be cut
down by a factor $\sim (a/2)^2$. Since the chiral restoration effect
is operative only above the ``flash temperature"~\cite{BR:PR}
$T_{flash}\sim 120$ MeV (and/or the ``flash density" $n_{flash}\lsim
2 n_0$), the dileptons emitted carrying the imprint of chiral
restoration -- at which the $\rho\pi\pi$ coupling will be near zero
due to the VM -- will be highly diluted, if not totally swamped, by
dileptons coming from near on-shell. Unless the probe pinpoints the
kinematic regime above the flash points, it is difficult to imagine
how chiral properties can be clearly sorted out.

In sum, our answer to the question posed above is the $negative$ at
the present stage of experimental as well as theoretical
development. A corollary to this is that the null result in the
analyses of \cite{rapp} -- that mundane strong hadronic interactions
can more or less explain the available dilepton data -- does not
necessarily invalidate the HLS/VM scenario as some people have
argued. It merely indicates that whatever pertinent signals there
may be are masked by the mundane processes that have nothing direct
to do -- though not in inconsistency -- with chiral symmetry. In
this regard, our conclusion is in agreement with that arrived at by
Dusling, Teaney and Zahed~\cite{zahed-na60}.
\subsection{The Wess-Zumino-term induced process}
The process (\ref{Metag}) is mediated by the Wess-Zumino term that
arises from the Chern-Simons action in five dimensions in
AdS/QCD~\cite{SS}. In AdS/QCD theory, this process is
vector-dominated as \be
 \omega\rightarrow \pi^0 \bar{\rho}\rightarrow
 \pi^0\gamma^*\label{omegadecay}
 \ee
where $\bar{\rho}=\rho, \rho^\prime, \cdots$. It is noteworthy that
the direct coupling $\omega\pi\gamma$ is totally absent.
Surprisingly even in HLS$_1$ theory of Harada and Yamawaki which
does not exclude this direct coupling on a fundamental ground, an
analysis~\cite{HY:PR} showed that the direct coupling in the
$\omega\rightarrow \pi^0\gamma$ decay is absent. We may therefore
ignore it.

For simplicity, we ignore hadronic loop corrections which do not
shift masses and take into account only the IBD (intrinsic
background dependent) effects. By putting asterisk for the HLS$_1$
constants with the intrinsic density dependence incorporated, we can
write the amplitude for (\ref{omegadecay}) as
 \be
T^{\mu\nu}[\omega(p,\mu),\pi^0, \gamma^* (k,\nu)]\sim \frac{eg^*
N_c}{8\pi^2 F_\pi^*}\epsilon^{\mu\nu\alpha\beta} p_\alpha k_\beta
\frac{{m_\rho^*}^2}{{m_\rho^*}^2 -q^2}.\label{tree}
 \ee
We of course need to include hadronic loop corrections (e.g., width
for the $\rho$ as well as for the $\omega$) to make a quantitative
estimate. However one can note that first of all it has no $a$
dependence, hence no suppression of the type present in the dilepton
vertices. Second, the near degeneracy of the $\rho$ and $\omega$ in
medium could have an interesting effect, modulo of course the
broadening widths associated with hadronic loops. Since
anomaly-mediated processes are in general much cleaner than normal
processes, it would be interesting to analyze the CBELSA/TAPS
experiment taking into account the anomaly structure present in the
HLS framework.
\section{Dense Matter and Half-Skyrmions}
Hadronic physics at high density relevant to the interior of compact
stars is poorly understood at present. In fact we don't know much
about how hadronic matter transforms into a quark matter. Lattice
QCD cannot yet access high enough density and analytic tools of QCD
cannot handle the nonperturbative regime at a density which is not
high enough for the color-flavor locking mechanism to dominate.
Given the paucity of model-independent tools, we can be allowed to
speculate on an intriguing and hitherto unexplored scenario for the
transition to quark matter at high density based on the
instanton/skyrmion structure in holographic QCD.

It has been observed in the study of dense baryonic matter at high
density in terms of the Skyrme model put on crystals~\cite{BYP} that
a skyrmion immersed in dense medium fractionizes at a certain
density $n_{meron}> n_0$ into two
half-skyrmions~\cite{halfskyrmions}. At that density, the quark
condensate $\la\bar{q}q\ra$ is found to vanish but the pion decay
constant remains non-zero. In terms of the condensate written as
 \be
\la\bar{q}q\ra=\la\chi{\Tr}(U+U^\dagger)\ra
 \ee
what happens at $n_{meron}$ can be understood by that
$\la{\Tr}(U+U^\dagger)\ra=0$ and $\chi\neq 0$. This means that
chiral symmetry is restored but there is a ``gap" characterized by
$\chi\sim f_\pi\neq 0$. This resembles the pseudogap phenomenon in
high -$T_c$ superconductivity~\cite{pseudogap}.

In the AdS/QCD picture with hidden local fields, the half-skyrmions
correspond to merons, i.e., half-instantons. One can think of what's
happening as the liberation of the merons that are confined inside
the instantons at normal density into two deconfined merons at
$n_{meron}$, the latter becoming the relevant degrees of freedom in
the ``pseudogap" phase . This process in the presence of hidden
gauge fields resembles what happens in the quantum deconfinement
phenomenon in condensed matter physics~\cite{QDP}.\footnote{
Briefly, in the phase transition between the magnetically ordered
N\'eel state and a paramagnetic valence-bond solid (VBS) state, it
is believed~\cite{QDP} that a skyrmion in (1+2) dimensions made out
of the spinon field $\hat{n}$ fractionizes into two half-skyrmions
(or ``merons") at the phase boundary. The phase transition involves
no order parameters but is characterized by new degrees of freedom,
i.e., merons, specific to the critical point. Here a crucial role is
played by the emergent $U(1)$ gauge degree of freedom present in the
spinon field $\hat{n}$ when the latter is written in the so-called
CP$^1$ parametrization $\hat{n}=z^\dagger \vec{\sigma}z$ which has a
redundancy -- here $U(1)$ -- in a close analogy to the chiral field
$U$ described above. The confinement-deconfinement process of the
merons turns out to be intimately connected to the instanton in the
$U(1)$ gauge field.} If what happens at the chiral transition in
dense baryonic matter is of a similar nature to what happens in the
magnetic N\'eel-VBS transition, then there can be a variety of
interesting physical phenomena associated with the topological
structure of the half-skyrmions. One important immediate consequence
is that there could be a strong deviation from Fermi liquid
structure. Thus one could expect the ``normal" matter from which the
transition to a Cooper-paired state at high density takes place to
be a non-Fermi liquid state. This would bring a major change to the
description of the phase transition to a color-superconducting phase
from the hadronic phase which has always been assumed to be in a
Fermi-liquid state~\cite{rajagopal}.
\section{Summary and Discussions}
We addressed the question of how hadron masses arise and how they
can be tweaked by temperature/density to disappear in the framework
of hidden local symmetry at low energy. We have argued that in order
to theoretically describe dropping vector-meson masses at high
temperature and density, hidden local symmetry is indispensable.
Furthermore hidden local symmetric fields arise naturally as
emergent gauge fields bottom-up in chiral field theory as well as as
low-energy degrees of freedom top-down in holographic QCD. In either
way, one expects to have an infinite tower of vector fields coupled
gauge invariantly to the Goldstone bosons. Baryons must emerge in
this theory as skyrmions in the infinite tower of vector mesons in
four dimensions encapsulated in instantons in five dimensions.

Although AdS/QCD seems to work well for chiral dynamics of low-lying
hadrons in the vacuum, in describing what happens to the hadrons
under extreme conditions where the vacuum change must take place for
which $1/\lambda$ and $1/N_c$ corrections are crucial, we are
compelled to limit ourselves to the lowest member of the tower,
namely, the $\rho$ and $\omega$ (and $a_1$ if needed). This
corresponds to the HLS$_1$ of Harada and Yamawaki which we interpret
as a truncated HLS theory with a Wilsonian matching to QCD at a
matching scale commensurate with the chiral scale. This theory has
the ``vector manifestation fixed point" at which hadron masses
vanish (in the chiral limit) in a way equivalent to Brown-Rho
scaling. We have argued that certain consequences of the VM fixed
point structure must be incorporated in the theoretical framework in
confronting heavy-ion experimental data and that in the absence of
such analyses,  the presently available experimental data can
neither validate nor invalidate the scenario associated with the
hidden local symmetry with the VM fixed point or Brown-Rho scaling.
Finally we have suggested that the structure of baryons in terms of
instantons or skyrmions embedded in an infinite tower of vector
mesons could bring a drastic change to the phase transition scenario
at high density.
\section*{Acknowledgments}
Most of the ideas presented in this paper have been developed in
collaboration with Gerry Brown who however should not be held
responsible for the more speculative parts of the discussions. This
work was supported in part by the KRF Grants KRF-2006-209-C00002.


%


\end{document}